extracted from feature maps at various scales. It has consistently demonstrated impressive performance in various tasks [3-5].

*B. Gossip Mutual Learning*

We introduced mutual learning [6] into Gossip Protocol [7] to enable the simultaneous training of both local and incoming models on the local data in a decentralized way. Model alignment was achieved by introducing an additional loss function based on the Kullback-Leibler Divergence (KLD) between the predictions of these two models. To emphasize the model agreement in the tumor area, we developed a regional KLD (rKLD) loss. For two models with predictions $P_1$ and $P_2$ correspondingly on voxels $x$, let $T$ represent the ground truth of segmentation, which is normally a manually identified tumor region, the rKLD loss with respect to $T$ is defined as:

$$rKLD(P_1, P_2|T) = \sum_{i,j,k} P_1(x_{ijk}) \log \frac{P_1(x_{ijk})}{P_2(x_{ijk})} t_{ijk} \quad (1)$$

where $\Sigma_{i,j,k}$ represents the sum over all voxel indices *(i, j, k)* for a 3D image, $t_{ijk} \in \{0,1\}$ is the **true class** of $x_{ijk}$, with $t_{ijk} = 1$ for tumor and $t_{ijk} = 0$ for background.

Likewise, let *T'* represents the predicted segmentation by $P_1$, we can also compute the rKLD loss with respect to *T'* as:

$$rKLD(P_1, P_2|T') = \sum_{i,j,k} P_1(x_{ijk}) \log \frac{P_1(x_{ijk})}{P_2(x_{ijk})} t'_{ijk} \quad (2)$$

where $t'_{ijk} \in \{0,1\}$ is the **predicted class** of $x_{ijk}$, with $t'_{ijk} = 1$ for tumor and $t'_{ijk} = 0$ for background.

The mixed rKLD loss can then be computed as:

$$rKLD(P_1, P_2|M) = \frac{rKLD(P_1, P_2|T) + rKLD(P_1, P_2|T')}{\sum_{i,j,k} t_{ijk} + \sum_{i,j,k} t'_{ijk}} \quad (3)$$

where *M* stands for the presence of both *T* and *T'*.

The general scheme of GML workflow is illustrated in Fig. 1. Let the *i-th* participating site be associated with local dataset $D_i$ and trainable model $W_i$. The major steps of GML (with matching indices marked in Fig. 1) include:

1. Randomly initialize all participating sites' local models, then train each model in parallel using its local dataset $D_i$ for a certain number of iterations, to allow "model warm-up".

*Abstract*—Federated Learning (FL) enables collaborative model training among medical centers without sharing private data. However, traditional FL risks on server failures and suboptimal performance on local data due to the nature of centralized model aggregation. To address these issues, we present Gossip Mutual Learning (GML), a decentralized framework that uses Gossip Protocol for direct peer-to-peer communication. In addition, GML encourages each site to optimize its local model through mutual learning to account for data variations among different sites. For the task of tumor segmentation using 146 cases from four clinical sites in BraTS 2021 dataset, we demonstrated GML outperformed local models and achieved similar performance as FedAvg with only 25% communication overhead.

*Clinical Relevance*— Automatic segmentation of brain tumors is an essential but challenging step for extracting quantitative imaging biomarkers for accurate tumor detection, diagnosis, prognosis, treatment planning and assessment. We proposed a novel learning schedule to build an effective AI model by learning from data across different sources without sacrificing patient privacy, which can be generalized to a broader clinical research and applications.

I. INTRODUCTION

Federated learning (FL) [1] enables collaboration among medical centers for training deep-learning models like tumor segmentation without sharing private data. However, conventional FL like FedAvg [1] relies on a central server for global model aggregation, making it vulnerable to server failure and less effective in handling data variations among centers. To overcome these limitations of FL, we proposed Gossip Mutual Learning (GML), a decentralized collaborative learning framework that employs Gossip Protocol for direct peer-to-peer communication and encourages each center to optimize its local model by leveraging useful information from peers through mutual learning.

II. METHODS

*A. Data and Segmentation Model*

We demonstrated our approach on the task of brain tumor segmentation using multi-parametric MRI data from four contributing sites in BraTS 2021 dataset [2], as shown in Table 1. The Scale Attention Network (SANet) [3], [4] was employed as the segmentation model. SANet leverages a dynamic scale attention mechanism to incorporate fine-grained details with high-level semantic information

* This research is supported by a research grant from Varian Medical Systems (Palo Alto, CA, USA), UL1TR001433 from the National Center for Advancing Translational Sciences, and R21EB030209 from the National Institute of Biomedical Imaging and Bioengineering of the National Institutes of Health, National Institutes of Health, USA.

Yading Yuan (corresponding author) is with the Columbia University Medical Center, New York, NY 100032 USA (e-mail: yading.yuan@columbia.edu).

Jingyun Chen is with the Columbia University Medical Center, New York, NY 100032 USA (e-mail: jc6171@cumc.columbia.edu).

2. Each participant has a probability to become the model receiver r at the $t$-th iteration, upon which another randomly selected site will become the model sender $s$, and will send its local model $W_s$ to the receiver site.

3. On the receiver site, the local model $W_r$ and incoming model $W_s$ are updated via regionalized mutual learning. The corresponding loss functions for $W_r$ and $W_s$ are defined as:

$$L(W_r|W_s, M_r) = (1-\lambda) * JD(P_r, M_r) + \lambda * rKLD(P_r, P_s|M_r) \quad (4)$$

$$L(W_s|W_r, M_r) = (1-\lambda) * JD(P_s, M_r) + \lambda * rKLD(P_s, P_r|M_r) \quad (5)$$

where $P_s$ and $P_r$ are the predictions of models $W_s$ and $W_r$ on data $D_r$ correspondingly, $M_r$ represents the true and predicted segmentations on data $D_r$, $JD()$ is the Jaccard distance between prediction and ground truth [8], and $rKLD()$ is the mixed rKLD loss defined in Equation (3). $\lambda$ represents the weight of rKLD loss in the total loss. We set $\lambda$ as 0.9.

4. Finally, the receiver's local model is replaced by the weighted average between the updated $W_r$ and $W_s$ models from step 3.

*C. Baselines and Evaluation*

We implemented the proposed GML method with PyTorch and compared its performance with three baseline methods: Pooled model (PM) that was trained with all the data from four sites; individual Model (IM) that was trained with data on its own data without exchange, and FedAvg [1]. We used averaged Dice Similarity Coefficient (DSC) as evaluation metric in our study. All the experiments were conducted on four Nvidia GTX 1080 TI GPUs. In this work, we randomly split the 4 sites into two pairs of sender and receiver at each iteration, with each pair performing one model exchange.

## III. RESULTS

First, we tested the **local performance** of GML and bassline methods on each site's own testing data (Table 2). The final global models were tested for both PM and FedAvg, while site-specific models were tested for both GML and IM. As shown in Table 2, GML achieved higher performance than IM on three of the four sites. Particularly on Site 2, GML achieved highest DSC than all three baseline methods. This indicates the advantage of personalized learning under decentralized framework. GML also showed comparable performance with FedAvg across all sites, but with 25% communication overhead.

Second, we tested the **global performance** of GML, FedAvg and PM on the combined testing cases from all four sites. Given that GML maintains a local model on each site, we adopted a bagging-like strategy to aggregate the outputs of these site-specific models when evaluating GML's overall performance against global models from Pooled training and FedAvg. As shown in Table 3, GML displayed comparable performance to both PM and FedAvg.

## IV. CONCLUSIONS

We proposed GML as a decentralized learning framework for automatic tumor segmentation on multi-parametric MRI. GML offers a robust alternative to centralized FL as it does not rely on a central server to aggregate models from different clients. Decentralized learning also enables efficient communications among participating sites as it avoids the needs of model aggregation and redistribution in FedAvg. Meanwhile, GML encourages the enhancement of local models, therefore may yield improved performance on the local data.

TABLE I.  NUMBERS OF TOTAL, TRAINING, VALIDATING AND TESTING CASES FOR THE PARTICIPATING SITES

|  | Site 1 | Site 2 | Site 3 | Site 4 |
|---|---|---|---|---|
| Total Cases | 47 | 34 | 30 | 35 |
| Training Cases | 32 | 23 | 21 | 23 |
| Validating Cases | 5 | 4 | 3 | 4 |
| Tetsing Cases | 10 | 7 | 6 | 8 |

TABLE II.  MEAN DSCs ON TESTING CASES FROM INVIDIUAL SITES. THE FINAL GLOBAL MODELS ARE TESTED FOR PM AND FEDAVG, WHILE SITE-SPECIFIC MODELS ARE TESTED FOR GML AND IM.

|  | Site 1 | Site 2 | Site 3 | Site 4 |
|---|---|---|---|---|
| PM | 0.9515 | 0.8941 | 0.8547 | 0.9537 |
| FedAvg | 0.9369 | 0.8894 | 0.8405 | 0.9425 |
| IM | 0.9227 | 0.8749 | 0.8410 | 0.9349 |
| GML | **0.9351** | **0.8945** | **0.8419** | **0.9315** |

TABLE III.  MEAN DSCs ON TESTING CASES FROM ALL SITES COMBINED. SITE-SPECIFIC GML MODELS ARE AGGREGATED FOR TESTING.

|  | Combined |
|---|---|
| PM | 0.9203 |
| FedAvg | 0.9095 |
| GML (essembled) | **0.9104** |

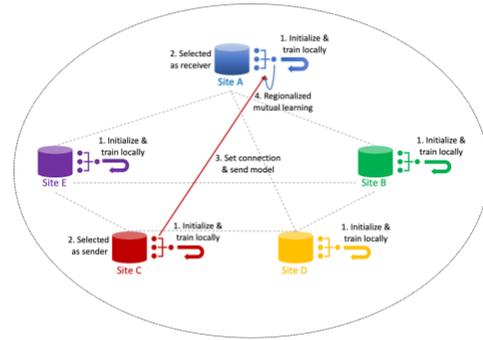

Figure 1. General scheme of GML. After initialization and warm-up, pairs of sender and receiver are randomly selected at each round. Then the sender transfers its model to the corresponding receiver. And the receiver performs regionalized mutual learning between incoming model and its local model.